\documentclass{article}

\usepackage{arxiv}

\usepackage[utf8]{inputenc} 
\usepackage[T1]{fontenc}    
\usepackage{hyperref}       
\usepackage{url}            
\usepackage{booktabs}       
\usepackage{amsfonts}       
\usepackage{nicefrac}       
\usepackage{microtype}      
\usepackage{xcolor}         

\usepackage{times}
\usepackage{epsfig}
\usepackage{graphicx}
\usepackage{amsmath}
\usepackage{amssymb}
\usepackage{subfig}

\newcommand{\scparin}{\phi_\mathrm{in}}
\newcommand{\scparout}{\phi_\mathrm{out}}
\newcommand{\tp}{\mathbf{p}}

\DeclareUnicodeCharacter{2206}{$\Delta$}

\title{Unsupervised learning of MRI tissue properties using MRI physics models}

\author{Divya Varadarajan \\
	Department of Radiology, \\
	Harvard Medical School, MA 02115 \\
	Massachusetts General Hospital, MA 02114\\
	\texttt{dvaradarajan@mgh.harvard.edu} \\
	\And
	Katherine L. Bouman \\
	Computing and Mathematical Sciences (CMS),\\
	Department of Electrical Engineering,\\
	Department of Astronomy, \\
	California Institute of Technology, CA 91125 \\
	\texttt{klbouman@caltech.edu} \\
	\AND
    Andre van der Kouwe\\
	Department of Radiology, \\
	Harvard Medical School, MA 02115 \\
	Massachusetts General Hospital, MA 02114\\
	\texttt{avanderkouwe@mgh.harvard.edu} \\
	\And
    Bruce Fischl\\
	Department of Radiology, \\
	Harvard Medical School, MA 02115 \\
	Massachusetts General Hospital, MA 02114\\
	\texttt{bfischl@mgh.harvard.edu} \\
	\And
    Adrian V. Dalca\\
	Department of Radiology, \\
	Harvard Medical School, MA 02115 \\
	Massachusetts General Hospital, MA 02114\\
	Electrical Engineering and Computer Science Department, \\
	Massachusetts Institute of Technology, MA 02139 \\
	\texttt{adalca@mgh.harvard.edu} \\
}

\renewcommand{\shorttitle}{\textit{arXiv} Template}

\hypersetup{
pdftitle={Unsupervised learning of MRI tissue properties using an MRI physics model},
pdfsubject={q-bio.NC, q-bio.QM},
pdfauthor={David S.~Hippocampus, Elias D.~Striatum},
pdfkeywords={First keyword, Second keyword, More},
}

\begin{document}
\maketitle

\begin{abstract}
	In neuroimaging, MRI tissue properties characterize underlying neurobiology, provide quantitative biomarkers for neurological disease detection and analysis, and can be used to synthesize arbitrary MRI contrasts. Estimating tissue properties from a single scan session using a protocol available on all clinical scanners promises to reduce scan time and cost, enable quantitative analysis in routine clinical scans and provide scan-independent biomarkers of disease. However, existing tissue properties estimation methods - most often $\mathbf{T_1}$ relaxation, $\mathbf{T_2^*}$ relaxation, and proton density (PD) - require data from multiple scan sessions and cannot estimate all properties from a single clinically available MRI protocol such as the multiecho MRI scan. In addition, the widespread use of non-standard acquisition parameters across clinical imaging sites require estimation methods that can generalize across varying scanner parameters. However, existing learning methods are acquisition protocol specific and cannot estimate from heterogenous clinical data from  different imaging sites. In this work we propose an unsupervised deep-learning strategy that employs MRI physics to estimate all three tissue properties from a single multiecho MRI scan session, and generalizes across varying acquisition parameters. The proposed strategy optimizes accurate synthesis of new MRI contrasts from estimated latent tissue properties, enabling unsupervised training, we also employ random acquisition parameters during training to achieve acquisition generalization. We provide the first demonstration of estimating all tissue properties from a single multiecho scan session. We demonstrate improved accuracy and generalizability for tissue property estimation and MRI synthesis. 
\end{abstract}

\keywords{quantitative MRI, relaxation parameters, proton density, multiecho MRI, parameter estimation}

\section{Introduction}
Magnetic resonance imaging (MRI) is a powerful modality for imaging anatomy~\cite{lerch2017}, function~\cite{bandettini2012,kundu2017,kwong1992}, metabolism~\cite{hyder2017}, and pathology~\cite{desikan2009,geraldes2018}. Biological tissue has a characteristic density and time constants~\cite{barnaal1968,bojorquez2017,bottomley1984}, referred to as tissue properties, that represent how the tissue responds to the magnetization environment of the scanner. These MRI tissue properties can thus quantitatively characterize the underlying microstructure and are often used as biomarkers to detect and study mechanisms of various diseases~\cite{andreasen1991,baudrexel2010,larsson1988} in a scan independent manner. In addition, combining tissue properties with physics-based forward models facilitates synthesis of MRI contrast that was not acquired, easing the need for long acquisitions, enabling more cost-effective and practical MRI~\cite{griswold2000,lustig2005,pruessmann1999}. MRI synthesis could also provide more data points to clinicians to make well-informed diagnosis related decisions.

Clinical neuroimaging data acquired routinely at hospitals and clinics contain a wealth of knowledge. These datasets are becoming available for large retrospective studies, promising to lead to better understanding of the neurobiology of many disease processes. One way to understand disease is to study tissue properties estimated from these data. 

Tissue property estimation is an ill-posed inverse problem that currently requires many MRI scans that are seldom acquired in routine clinical examination due to constraints on total scan time. In addition, acquisition protocol are not standardized across existing clinical imaging sites, leading to large heterogeneous datasets. Estimating tissue properties in such data requires an estimation method to generalize across varying scanner settings. In this paper we present the first generalizable unsupervised learning framework that can estimate the three tissue properties (that are the sources of contrast in a standard structural MRI acquisition~\cite{bloch1946}), $\mathbf{T_1}$ relaxation time, $\mathbf{T_2^*}$ relaxation time and proton density (PD) from a single fast low angle shot (FLASH) multiecho MRI scan session with arbitrary scanner settings. The FLASH multiecho MRI protocol is a low scan-time clinically relevant protocol that is readily available on all clinical scanners. Our approach thus enables tissue property estimation from routine clinical data, and also reduces scan time by enabling the synthesis of unseen MRI contrasts~\cite{bloch1946,bobman1985}.

\begin{figure*}[t!]
\centering
\includegraphics[trim=50 130 50 50, clip, width=0.95\textwidth]{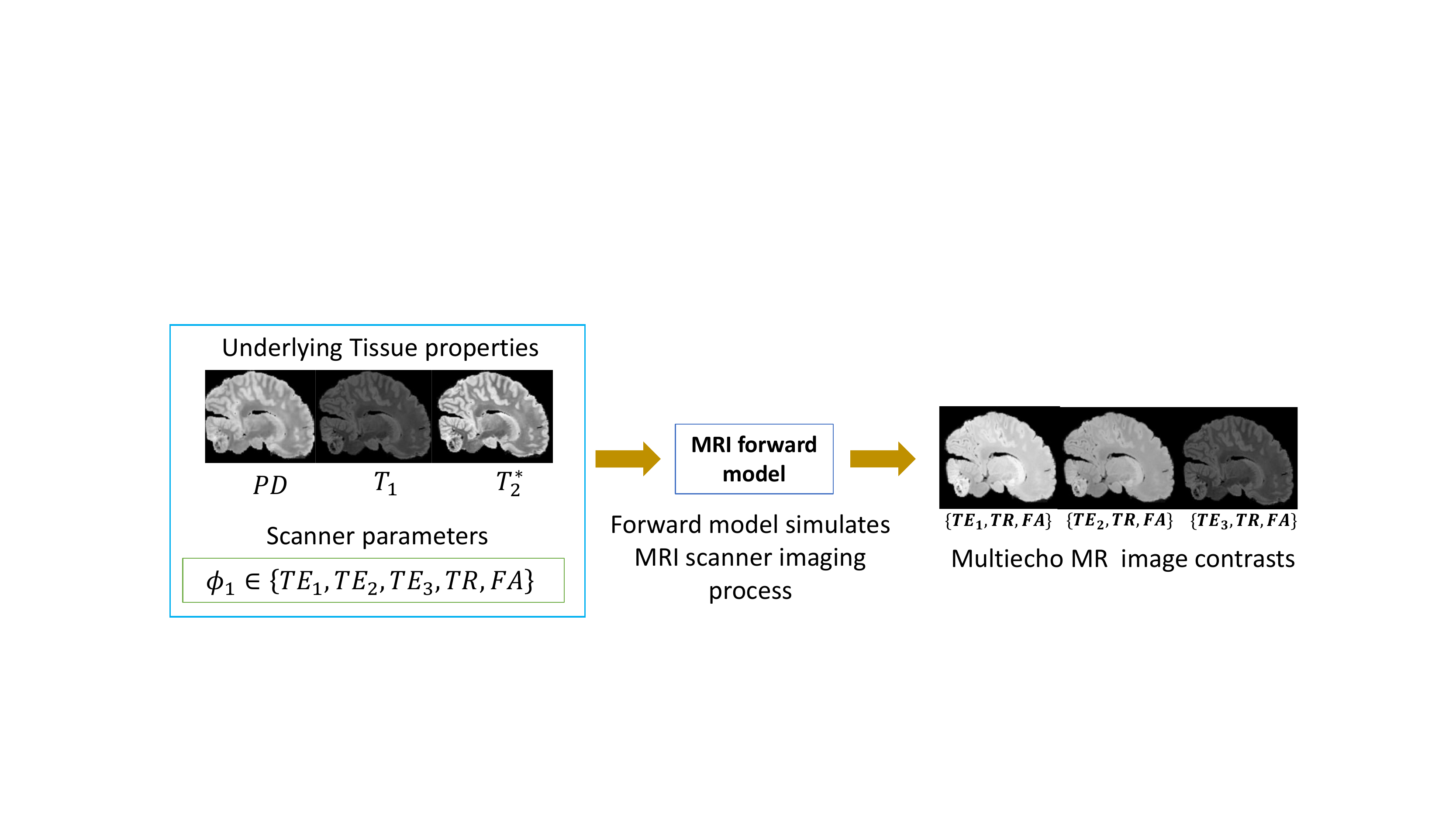}
\caption{The imaging process of an FLASH multiecho MRI scan session generates multiple MRI image contrasts. Each MRI contrast is associated with a unique set of scanner parameters consisting of repetition time (TR), echo time (TE) and flip angle (FA). The scanner parameters for the entire multiecho scan session, $\phi_1$ will contain all the scanner parameters of the the individual contrasts. \vspace{-0.5cm}}
\label{fig:mri_process}
\end{figure*}
MRI is based on the interaction of the biological tissue being imaged with the magnetization created by the MRI scanner. The imaging process of the scanner is parameterized by various timing parameters and scanner coil voltages, which we refer to as scanner parameters. The MR image intensity is generated due to the interaction of the tissue properties with scanner parameters within an MR physics process. Figure~\ref{fig:mri_process} illustrates this imaging process for a standard single FLASH multiecho MRI scan session. The scanner parameters, echo times (TE), repetition time (TR) and flip angle (FA), along with the tissue properties, $\mathbf{T_1}$, $\mathbf{T_2^*}$ and $\mathbf{PD}$, interact within the physics based MRI forward model to generate multiple MR image contrasts, with each contrast associated with a unique TE.

Existing estimation methods~\cite{ben-eliezer2015,fischl2004,hahn1949,hong2019,huang2014,kim2021,rieger2017,torrey1949} rely on measurements from multiple such multiecho scan sessions~\cite{deoni2003,fischl2004} to solve the inverse problem for estimating all three tissue properties. Estimation of all three tissue properties using data from a \textit{single} multiecho scan session, however, is still an open problem due to the highly ill-posed nature of the inverse problem in data starved settings. 
In addition, existing deep learning methods are trained for a fixed set of scanner parameters~\cite{cai2018,cohen2018,fang2019,liu2019,liu2021}, which greatly limits their deployment in heterogeneous clinical datasets where a variety of scan parameters are used, and are unable to estimate all three tissue properties from a \textit{single} multiecho scan session. Most learning methods are also supervised and require the tissue properties be provided during training, which greatly limits their applicability in real scenarios. 

Tissue properties also facilitate synthesizing arbitrary MRI contrasts that provide clinicians with many images with varying contrast to aid their decision making. Physics-based synthesis relies on first estimating the tissue properties from a few acquired MRI scans of a given subject and then using these to synthesize new scans through a well-characterized imaging process of this type of MRI~\cite{bloch1946}. The accuracy of synthesis thus relies heavily on the accuracy of the tissue property estimates.

In this paper we develop an unsupervised learning-based method that uses MRI physics-based forward models to accurately estimate tissue properties and synthesize unseen MRI contrasts.We employ a new training strategy that varies acquisition parameters, enabling the network to generalize to heterogeneous data from multiple imaging sites that use different acquisition parameters. We demonstrate the ability to estimate all three tissue properties from a single multiecho scan session to substantially outperform existing methods.

\section{Related Work}
In this section we discuss past work in acquisition and estimation strategies for tissue properties, and deep learning-based methods to perform parameter estimation.  

\subsection{Classical methods}
\label{sec:clopt}

Tissue property estimation methods solve an inverse problem for an MRI physics based forward model from multiple noisy measurements of the MR image contrasts that depend on the tissue property ~\cite{hahn1949,torrey1949}. Advancement in both MRI acquisition to acquire multiple parameter contrast data efficiently~\cite{griswold2000,lustig2005,pruessmann1999} and in estimation strategies~\cite{ben-eliezer2015,fischl2004,hahn1949,hong2019,huang2014,kim2021,rieger2017,torrey1949} to accurately predict the tissue properties have propelled the field of quantitative MRI.

These multiple indirect measurements of tissue properties are made by generating MRI contrast with varying scanner parameters. For example, MRI acquisitions with multiple parameter measurements include multiecho MRI where TE is varied to enable $\mathbf{T_2}$ or $\mathbf{T_2^*}$ estimation (Figure 1) , variable flip angle MRI~\cite{wang1987} where FA is varied for $\mathbf{T_1}$ estimation, multiple inversion-recovery-prepared MRI where the inversion time (TI) is varied for $\mathbf{T_1}$ estimation~\cite{hahn1949,messroghli2004} and MR fingerprinting~\cite{ma2013} where multiple scan parameters ($TR$, $TE$ and $FA$) are randomly varied for joint estimation of $\mathbf{T_1}$, $T_2$, $\mathbf{T_2^*}$ and $PD$. 

Previous methods employ dictionary-based optimization where the dictionary samples the MR imaging model~\cite{ben-eliezer2015,hong2019,huang2014,kim2021,rieger2017} or numerical fitting techniques that use iterative optimization to minimize the nonlinear cost~\cite{block2009,deoni2003,fischl2004}. Forward models have been made more robust by incorporating the physics-based models of imaging artifacts~\cite{stikov2015,sijbers1999,venkatesan1998}. Methods that decrease scan time of each contrast by estimating parameters from undersampled MRI data use iterative optimization~\cite{sumpf2011}, sparsity constraints~\cite{lingala2011,velikina2013}, low rank constraints~\cite{zhang2015}, and compressed sensing algorithms~\cite{doneva2010} to reconstruct the signal from undersampled data and estimate tissue properties jointly.

\subsection{Deep learning based methods}
Supervised~\cite{cai2018,cohen2018,fang2019} and semi-supervised~\cite{liu2019} deep learning based methods train a neural network to estimate $\mathbf{T_1}$ and $T_2$ relaxation from complex valued or magnitude MRI data, using images of different biological tissue. The methods exhibit comparable accuracy to classical approaches, higher robustness to system imperfections, and fast execution time~\cite{feng2020}. However, supervised and semi-supervised methods require knowledge of the true tissue properties during training, which is impractical for many applications, and do not generalize well to MRI contrasts that the network has not seen during training. 

A recently proposed unsupervised deep learning method estimates $\mathbf{T_1}$ or $\mathbf{T_2})$ by training a network to reconstruct the input to the network~\cite{liu2021}. As we show in our experiments, since such methods only learn to synthesize the given input, they cannot synthesize other contrasts accurately. In addition, all existing deep learning methods assume a specific acquisition protocol and require re-training with new training data when adapting to a different acquisition protocol.

\begin{figure*}[t!]
\centering
\includegraphics[trim=50 205 50 30, clip, width=1\textwidth]{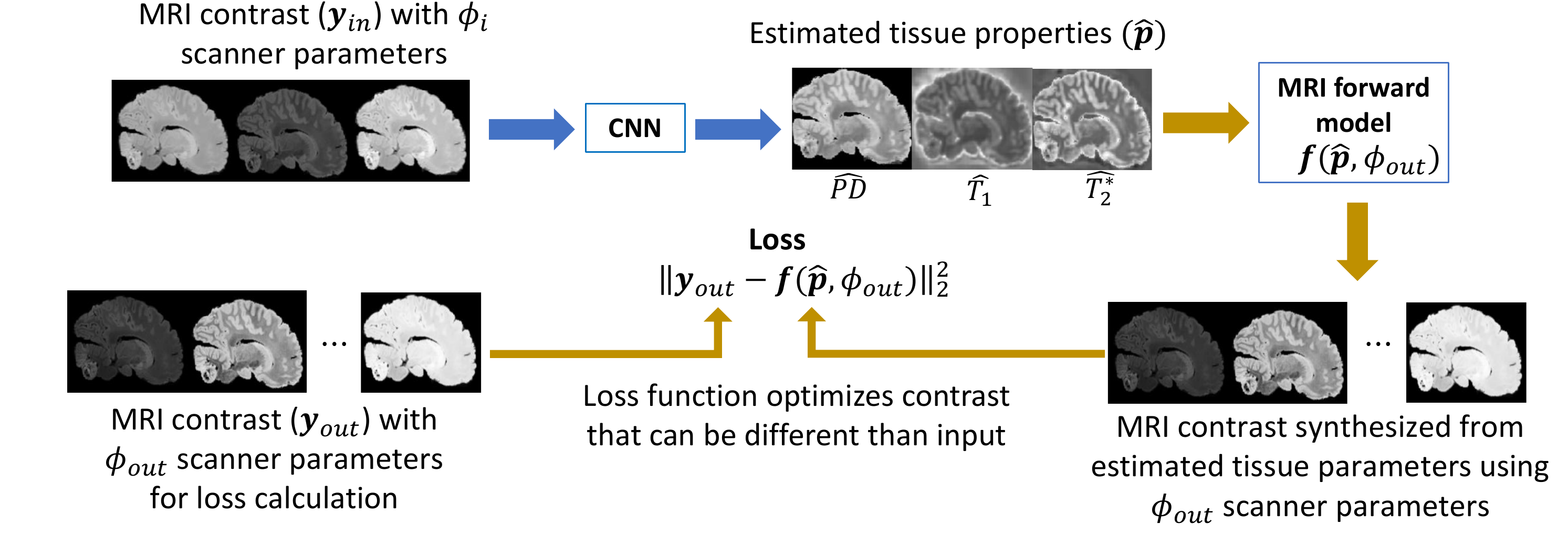}
\caption{Proposed model to synthesize arbitrary FLASH MRI contrasts using a CNN and a FLASH forward model from three input image contrasts. As a consequence of using the FLASH model, the output of the CNN can be interpreted as estimates of the tissue properties (T1,T2* and $\mathbf{PD}$).\vspace{-0.4cm}}
\label{fig:training}
\end{figure*}
 \section{Method}
In this section we setup the synthesis and tissue property estimation problem and explain the proposed method. We choose to demonstrate our method with FLASH multiecho MRI acquisition protocol~\cite{frahm1986}, a widely used MRI modality because of its availability in all clinical MRI scanners, ease of forward modeling~\cite{frahm1986} and its dependence on tissue properties that have clinical significance~\cite{andreasen1991, baudrexel2010,larsson1988}. 
Our proposed strategy generalizes to other acquisition protocols that have an appropriate forward model. 

\subsection{Background}
We first provide background on how forward models are used to synthesize MRI scans, and describe the physics based forward model for FLASH MRI acquisition.

\subsubsection*{Synthesis of MRI Contrast}
In the image acquisition community, MRI synthesis refers to the process of simulating other MRI contrasts from a set of acquired MRI scans. We first estimate the tissue properties~$\tp$ from acquired scans $\mathbf{y}$ of a subject,
\begin{equation}
\hat{\tp} = \arg \min_\tp ||h(\tp,\scparin) - \mathbf{y}||^2_2, \label{eq:param}
\end{equation}
where~${\scparin}$ are the scanner parameters of acquired MRI data,~$h(\cdot, \cdot)$ is a function capturing the MR physics forward model of the acquired scans based on tissue and scanner parameters. The estimated tissue properties~$\hat{\tp}$ are then incorporated into the forward model of the MRI contrast to be synthesized,
\begin{equation}
\mathbf{\hat{x}} = f(\hat{\tp},\scparout), \label{eq:synth}
\end{equation}
where $f(\cdot,\cdot)$ represents the MR physics based forward model of the modality to be synthesized,~$\scparout$ are the synthesis scanner parameters and~$\mathbf{\hat{x}}$ is the synthesized MRI. In this paper we use FLASH steady state acquisition for both forward models~$h(\cdot,\cdot)$ and $f(\cdot,\cdot)$.

\subsubsection*{FLASH MRI Contrast}
FLASH MRI is an imaging sequence that generates images that depend on three tissue properties: $\mathbf{T_1}$ relaxation time,~$\mathbf{T_2^*}$ relaxation time, and proton density ($\mathbf{PD}$)~\cite{bloch1946}. The dependence on tissue properties~$\tp \in \{\mathbf{T_1}, \mathbf{T_2^*}, \mathbf{PD}\}$ and scanner parameters~$\mathbf{\phi} \in \{TR, TE, \alpha\}$ is captured by a forward model derived from the Bloch equation~\cite{bloch1946,frahm1986}:
\begin{align}
     \mathbf{y} = f(\tp,\phi) = \mathbf{PD} \cdot sin(\alpha)  \exp(-TE/\mathbf{T_2^*}) \frac{(1 - E_1)}{(1 - \cos(\alpha)*E_1)},
      \label{eq:flash}
\end{align}
where~$E_1 = \exp(-TR/\mathbf{T_1})$, TR is the repetition time, $\alpha$ is flip angle (FA), TE is the echo time, and~$\mathbf{y}$ is the MRI intensity.

A multiecho scan session constitutes of acquiring image contrasts~$\mathbf{y}$ at multiple TE values to generate contrasts based on eqn,~\eqref{eq:flash}. Existing methods use multiecho session to estimate $T_2^*$. Similarly, based on eqn.~\eqref{eq:flash}, $T_1$ is estimated from scan sessions containing multiple FA values. Since the multiecho scan session uses one value of FA, it does not contain multiple measurements of $T_1$, making  $T_1$ estimation from a multiecho session highly illposed and challenging. Therefore existing methods acquire a lot more multiecho scan sessions with multiple FA values to estimate all three tissue properties, which is costly  (at least three times the time of a \textit{single} multiecho scan session). In contrast, we propose a method to estimate all three tissue properties, including $\mathbf{T_1}$ from  a single multiecho scan session, opening up many possibilities.

\subsection{Learning Formulation}
We propose an unsupervised learning framework that estimates tissue property maps~$\tp$ from arbitrary MRI contrasts. We assume a dataset where each item is a set of MR images~$\mathcal{Y} = \{\mathbf{y}\}$ of the same anatomy, and hence same unknown tissue properties~$\tp$, acquired using known varying scanner parameters~$\scparin$. Such acquisitions are common in clinical and research scenarios. 

We let function~$g_\theta(\mathcal{Y}) = \tp$ with parameters~$\theta$ map a set of input MR images~$\mathcal{Y} = \{\mathbf{y}\}$ to tissue properties~$\tp$. We propose an \textit{unsupervised} training strategy driven by the idea that estimated tissue properties should be able to synthesize new contrasts accurately using a physics based forward model. 
The loss function optimizes the ability to synthesize new MRI contrasts with the estimated tissue properties~$\hat{\tp}$ using a forward model~$f(\cdot, \cdot)$: 
\begin{equation}
 \mathcal{L}(\theta; \mathcal{Y}) =  \mathbb{E}_{\mathbf{y}_\mathrm{in} \in \mathcal{Y}, \scparout \in T}  \| f(\hat{\tp},\scparout) - \mathbf{y}_\mathrm{out} \|_2^2, 
\end{equation}
where~$\hat{\tp} = g_\theta(\mathbf{y}_\mathrm{in})$, $\scparout$ are the output (synthesis) scanner parameters with corresponding~$\mathbf{y}_\mathrm{out}$ MRI images, $T$ is the space of scanner parameters and $\mathbb{E}$ is the expectation operator. In our experiments, we use the FLASH forward model.

\textbf{Training.} We use a variety of scans~$\mathbf{y}_\mathrm{in}$, and~$\mathbf{y}_\mathrm{out}$, obtained using different scanner parameters as is common in clinical scenarios to train. We provide details of the scan parameter values that were used to generate the heterogeneous training data in Table 1. We hypothesize that this will help yield a network that generalizes well to a wide array of data. Figure~\ref{fig:training} illustrates a summary of the proposed strategy.

\textbf{Network architecture and implementation details.}
We use a U-Net architecture for the function~$g_\theta(\cdot)$~\cite{ronneberger2015} in our experiments, which takes 3 input FLASH MRI images and estimates 3 tissue properties. The encoder of the U-Net comprises of 6 blocks of 2D convolutions, ELU activation and max pooling that halved the resolution. The decoder of the U-Net comprises of 5 blocks of upsampling that doubled the resolution, 2D convolutions and ELU activation layer, followed by a last 2D convolution layer and ReLU activation layer. The number of filters and convolution kernel size are set to 64 and $3 \times 3$ respectively. The output of the U-net along with set of output scan parameters $\phi_{out}$ is passed to a FLASH MRI forward model layer that implements the forward model from Eq. (3). The forward model layer generates output FLASH MRI contrasts corresponding to $\phi_{out}$ scanner parameters. To avoid high GPU memory requirements when working with multi-contrast MRI we work with single slices at each training iteration. We implemented all models and the FLASH MRI forward model using Tensorflow~\cite{abadi2015}. We train the network with a single multiecho scan session input with 3 MR image contrasts, batch  size  of 6 slices and a learning rate of 0.001. All training was performed on an NVIDIA Quadro GV100 GPU with 32 GB memory and the proposed network took 39.5 hours to train.

\section{Experiments and Results}
We analyze the ability of the proposed method to estimate tissue property maps from MR image contrasts of a single multiecho input scan session, to generalize across multiple acquisition parameters and to synthesize arbitrary MRI contrasts.

\textbf{Datasets.} 
Our goal is to simulate the scenario often observed in practice, where a variety of sessions are acquired for different subjects using different scan parameters, leading to large sets of heterogeneous groups of scans. 
\def\arraystretch{0.5}%
\begin{figure}[h!]
\centering
\includegraphics[trim=5 200 0 110, clip, width=1\textwidth]{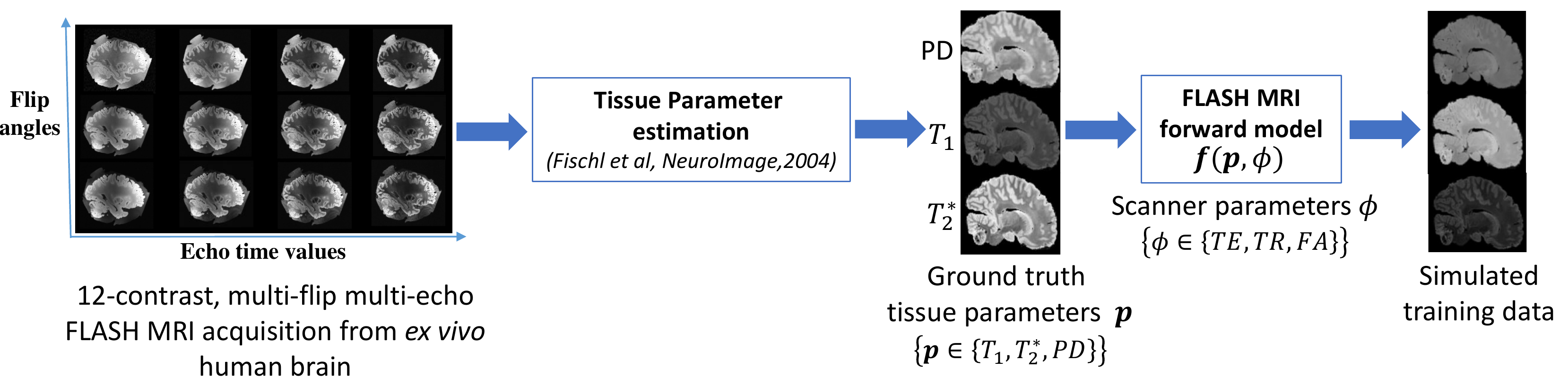}
\caption{Training Data simulation steps. We estimated tissue parameters~$\mathbf{p}$ from a large 12-contrast FLASH MRI acquisition containing multiple flip angles (FA) and echo time (TE) scanner parameters for 22 different \textit{ex vivo} human brain specimens using a dictionary-based method~\cite{fischl2004}. Estimated tissue parameters were used as ground truth to simulate training data. \vspace{-0.5cm}}
\label{fig:tr_sim} 
\end{figure}

To achieve this, we first acquired 22 \textit{ex vivo} human brain hemispheres, each with three flip angles in three separate scan sessions, substantially more than is available per subject. We then used a dictionary based method~\cite{fischl2004} to estimate the tissue property maps, $\mathbf{T_1}$, $\mathbf{T_2^*}$ and $\mathbf{PD}$, from this 3-flip angle, 4-echo FLASH MRI data. The acquired dataset had 1 mm isotropic  resolution, TE = [5 11 18 25] $ms$, TR = 37 $ms$, and FA ($\alpha$) = [10, 20, 30] degrees. Finally, using these tissue property maps, which we treat as ground truth for evaluation, we simulated a heterogeneous scan training dataset containing 3 FLASH MRI image contrast (~$\mathbf{y}_\mathrm{in}$) and randomly varying input scanner parameters~$\phi_{in}$ using the FLASH MRI forward model (Eq. ~\eqref{eq:flash}). We also simulated an additional set of 10 MR image output contrasts (~$\mathbf{y}_\mathrm{out}$) for each element in the training dataset to be used for loss optimization. We varied the scanner parameters in our training data to simulate the environment equivalent to heterogeneous clinical data across clinical sites, where it is common for each site to use a different set of scanner parameters. Table 1. summarizes the values of $\phi_{in}$ and $\phi_{out}$ used to generate our training dataset. We illustrate these steps taken to simulate our training data in Figure~\ref{fig:tr_sim}. 

\begin{table}
\centering
\begin{tabular}{||c||c|c||}
     \hline
    Network
    &\begin{tabular}{c}
          Randomly sample\\
          TE $\in$ 5 - 80 ms\\
          TR $\in$ 30 - 100 ms\\
          FA $\in$ 5 - 80 degrees\\
     \end{tabular}
     
    & \begin{tabular}{c}
    Fixed acquisition:\\TE=[7 15 25] ms,\\ TR=[37, 37, 37] ms,\\  FA=[20, 20, 20] degrees\\
    \end{tabular} \\
    \hline
    \hline
    Multi-acquisition network (proposed) & $\phi_{in}$  and $\phi_{out}$ & \\
    \hline
    Fixed acquisition network (baseline) & & $\phi_{in}$  and $\phi_{out}$ \\
    \hline
    Synthesis loss network &$\phi_{out}$  &  $\phi_{in}$ \\
     \hline
\end{tabular}
\caption{Training dataset parameters for proposed and baseline networks. \vspace{-0.5cm}}
\end{table}
\textbf{Test data setup and evaluation metric.}
We separate 20\% of the \textit{ex vivo} MRI volumes as held-out test data for performance evaluation. The corresponding tissue property maps that were estimated using procedure described above for these volumes were used as gold standard to simulate 3-echo test data and to evaluate accuracy of network estimated tissue properties. We simulated 1000 different test 3-echo MR images with $\scparin$ scanner parameters of each experiment and and additional 10 FLASH MR image contrast with $\scparout$ scanner parameters from the gold standard tissue properties dataset. The output contrast images corresponding to $\scparout$ were used as gold standard to test synthesis accuracy of the network. The input and output scanner parameters for each experiment is summarized in Table 2.

We used the mean absolute error (MAE) between the gold standard and the estimations from test data to evaluate both tissue properties estimation accuracy and synthesis performance of the proposed method. We also used absolute difference maps to show the spatial distribution of errors within an image slice.

\textbf{Baseline methods.}
To the best of our knowledge, existing methods cannot estimate all the three tissue properties from a single multiecho scan session. Specifically, they cannot estimate $\mathbf{T_1}$ and $\mathbf{PD}$, because to estimate these they normally require multiple MRI contrasts with different flip angle scanner parameter, which a single multiecho session keeps constant. Therefore, to best gain insight into the properties of the proposed method, we instead analyze ablated versions of the proposed model, which we trained on three-echo FLASH MRI with predetermined input scanner parameters $\mathbf{\phi}_{in}$. Details of the input and output scan parameters for this fixed acquisition network, which we refer to as baseline, is provided in Table 1. FA of 20 was chosen for the baseline because it maximizes the signal to noise ratio of our \textit{ex vivo} MRI scans. Baseline TR and TE were matched to be in close range of the acquisition protocol of the \textit{ex vivo} datasets that we used for evaluation, providing the ablation methods (or baseline) with the best scenario. In addition, the output scans~$\mathbf{y}_\mathrm{out}$ used to compute the loss function during training were the same as the input scans, as is standard in existing deep learning methods~\cite{feng2020}. This fixed acquisition based baseline network is also a generalization of previous learning methods that have tackled the problem of MRI quantitative estimation but only estimate $\mathbf{T_2^*}$ and/or $\mathbf{PD}$ from a single multiecho scan session.

We also trained another ablation network, referred to as the synthesis loss network, trained on the same predetermined input scanner parameters $\mathbf{\phi}_{in}$ as the baseline. However, it optimized output scans~$\mathbf{y}_\mathrm{out}$ that could be different from the input. Table 1 provides the details of the scan parameters. The synthesis loss network was used to test our hypothesis that the accuracy of MRI synthesis increases when networks optimize for MRI contrasts different from input contrasts during training.
\begin{table}
\centering
\begin{tabular}{||c||c|c||}
     \hline
     Experiment
     & Input $\phi_{in}$
     & Output $\phi_{out}$\\
     \hline
    \hline
    \begin{tabular}{c}
     Exp 1. Tissue property \\ estimation test,\\ Exp 3. MRI synthesis test,\\  Exp 4. Loss function test\\
    \end{tabular}
    & \begin{tabular}{c}
    TE=[7 15 25] ms,\\ TR=[37, 37, 37] ms,\\  FA=[20, 20, 20] deg\\
    \end{tabular}
    &\begin{tabular}{c}
          Randomly sample\\
          TE $\in$ 5 - 80 ms\\
          TR $\in$ 30 - 100 ms\\
          FA $\in$ 5 - 80 degrees\\
     \end{tabular}\\
     \hline
   \begin{tabular}{c}
   Exp 2. Acquisition \\ generalizability test
   \end{tabular}
    & \begin{tabular}{c}
    TE=[7 15 25] ms,\\ TR=[37, 37, 37] ms,\\ randomly sample \\FA $\in$ 0 - 40 degrees \\
    \end{tabular}
        &\begin{tabular}{c}
          Randomly sample\\
          TE $\in$ 5 - 80 ms\\
          TR $\in$ 30 - 100 ms\\
          FA $\in$ 5 - 80 degrees\\
     \end{tabular}\\
        \hline
\end{tabular}
\caption{Test scan parameters for each experiment setup. \vspace{-0.6cm}}
\end{table}

\subsection{Experiment 1: Tissue property estimation from a single multiecho scan session}

We first evaluated the ability to estimate all three tissue properties from a single multiecho scan session containing three MRI contrasts, which has not previously been done.  We qualitatively analyzed the estimated tissue properties as well as used the gold standard~$p$ to quantify MAE across 1000 test slices. The input scanner parameters~$\scparin$ for the test data was the same as that used to train the baseline method, providing the baseline with the best possible scenario. Essentially, this tests how generalizable the proposed method is in setting that the baseline is trained for directly. 

\noindent \textbf{Results}.
\begin{figure}[t]
\centering
\includegraphics[trim=20 60 0 10, clip, width=0.95\textwidth]{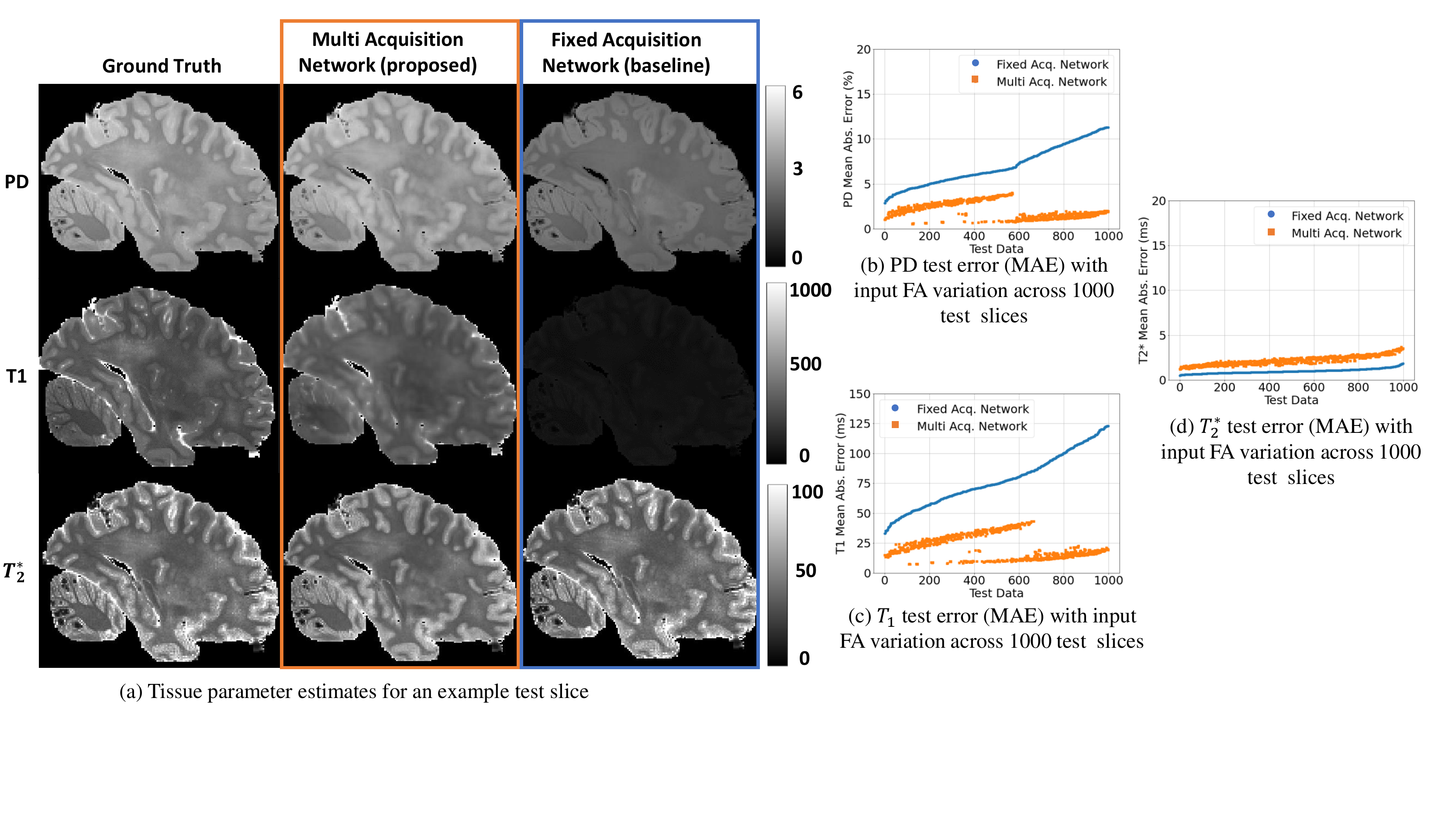}
\vspace{-0.5cm}
\caption{Tissue property Estimation. Subfigure (a) shows the ground truth image and tissue property image estimated by the proposed Multi-acquisition network and the baseline fixed acquisition network for an example test slice. Figures~\ref{fig:pmap_err}b-d plot MAE for each of the tissue property over 1000 test image slices. The plots are ordered based on increasing test errors of the baseline method.
\vspace{-0.3cm}}
 \label{fig:pmap_err} 
\end{figure}

\begin{figure}[h!]
\centering
\includegraphics[trim=300 0 300 0, clip, width=0.8\textwidth]{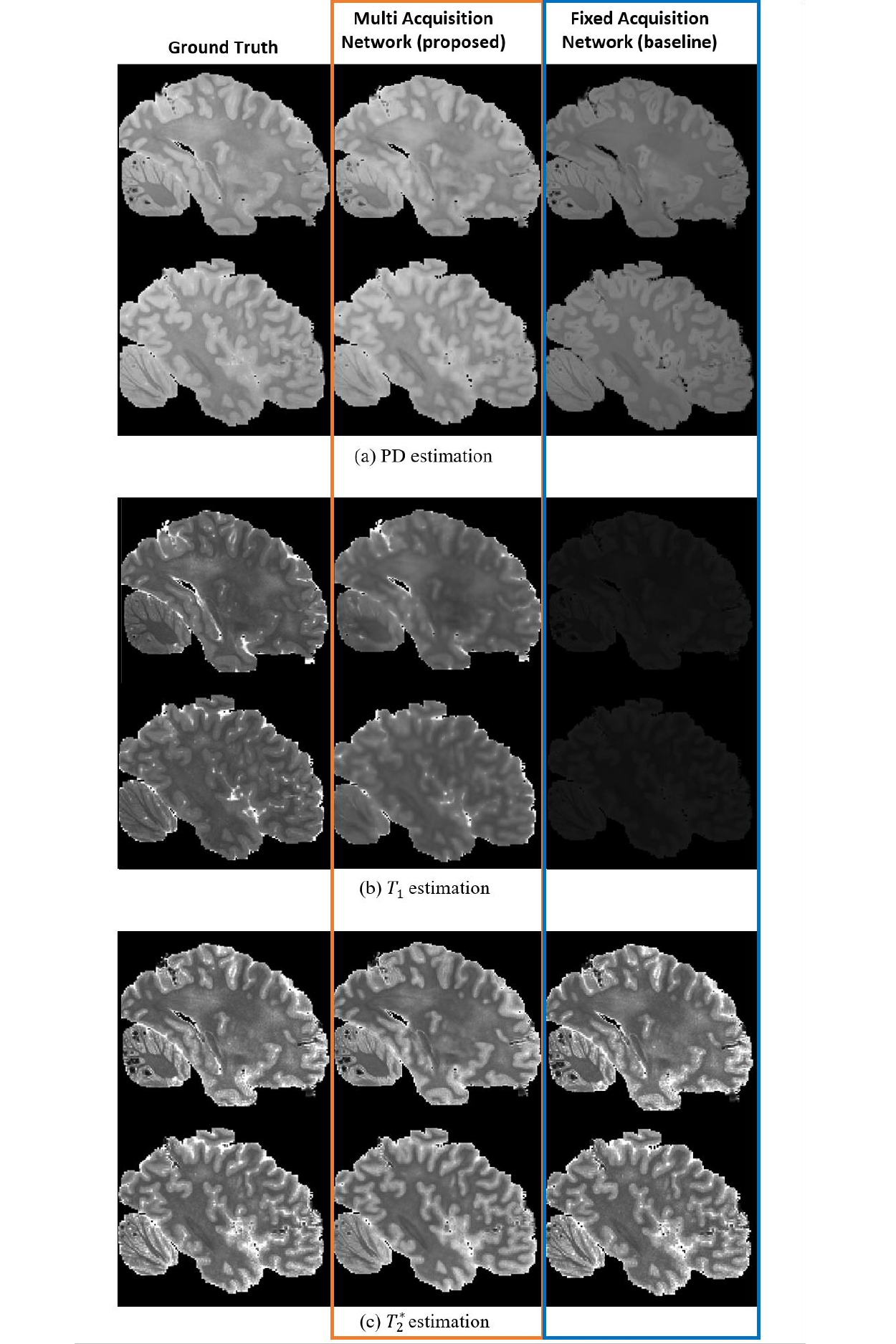}
\vspace{-0.2cm}
\caption{Additional example of tissue properties ($\mathbf{PD}$, $\mathbf{T_1}$, $\mathbf{T_2^*}$) estimated for two more test slices of experiment 1 where we estimated the tissue properties from 3-echo test data. The input scan parameters of test data was set same as the values used to train the baseline to give the baseline network an advantage. The proposed method estimates tissue properties $\mathbf{PD}$ and $\mathbf{T_1}$ substantially more accurately than baseline. $\mathbf{T_2^*}$ estimations are comparable across both proposed and baseline methods.\vspace{-0.5cm}}
\label{fig:pmap_err2} 
\end{figure}

Fig.~\ref{fig:pmap_err}a. shows the ground truth and estimated tissue properties for an example test image. Accurate $\mathbf{T_1}$ estimation for the baseline (ablation) would require data from multiple flip angles. As expected, since the 3-echo contrasts share the same flip angle, the baseline method estimates~$\mathbf{T_1}$  and~$PD$ incorrectly, while we observe that both parameters match the ground truth well for the proposed method. Training with multiple scanner parameters enabled estimation of~$\mathbf{T_1}$ from a multiecho scan session for the first time. Figs.~\ref{fig:pmap_err}b-d show the MAE for the three tissue parameters across 1000 test slices. The proposed method results in lower MAE for~$\mathbf{T_1}$ and~$PD$ estimates across 1000 test slices, with improvements by large margins of 20 $ms$ - 105 $ms$ and 3\% - 10\% respectively. For the $\mathbf{T_2^*}$ property estimate, the baselines performs slightly better (with error difference of 2 milliseconds ($ms$) or less), but this is in a regime where both methods perform extremely well. Additional examples in Fig.~\ref{fig:pmap_err2} are consistent and in agreement with our analysis.

The proposed method therefore estimates all three properties jointly from a single scan with large improvements in accuracy of $\mathbf{T_1}$ and $PD$ estimations and comparable performance with $\mathbf{T_2^*}$. This result is the first demonstration of~$\mathbf{T_1}$ estimation from a single multiecho scan session, an important first step for clinical applications where this holds promise to enable multidimensional quantitative analysis for existing datasets and clinical protocols that contain a  multiecho scan session.

\subsection{Experiment 2: Generalizability across acquisition parameters}

\begin{figure*}[bt!]
\centering
\includegraphics[trim=50 50 50 50, clip, width=0.95\textwidth]{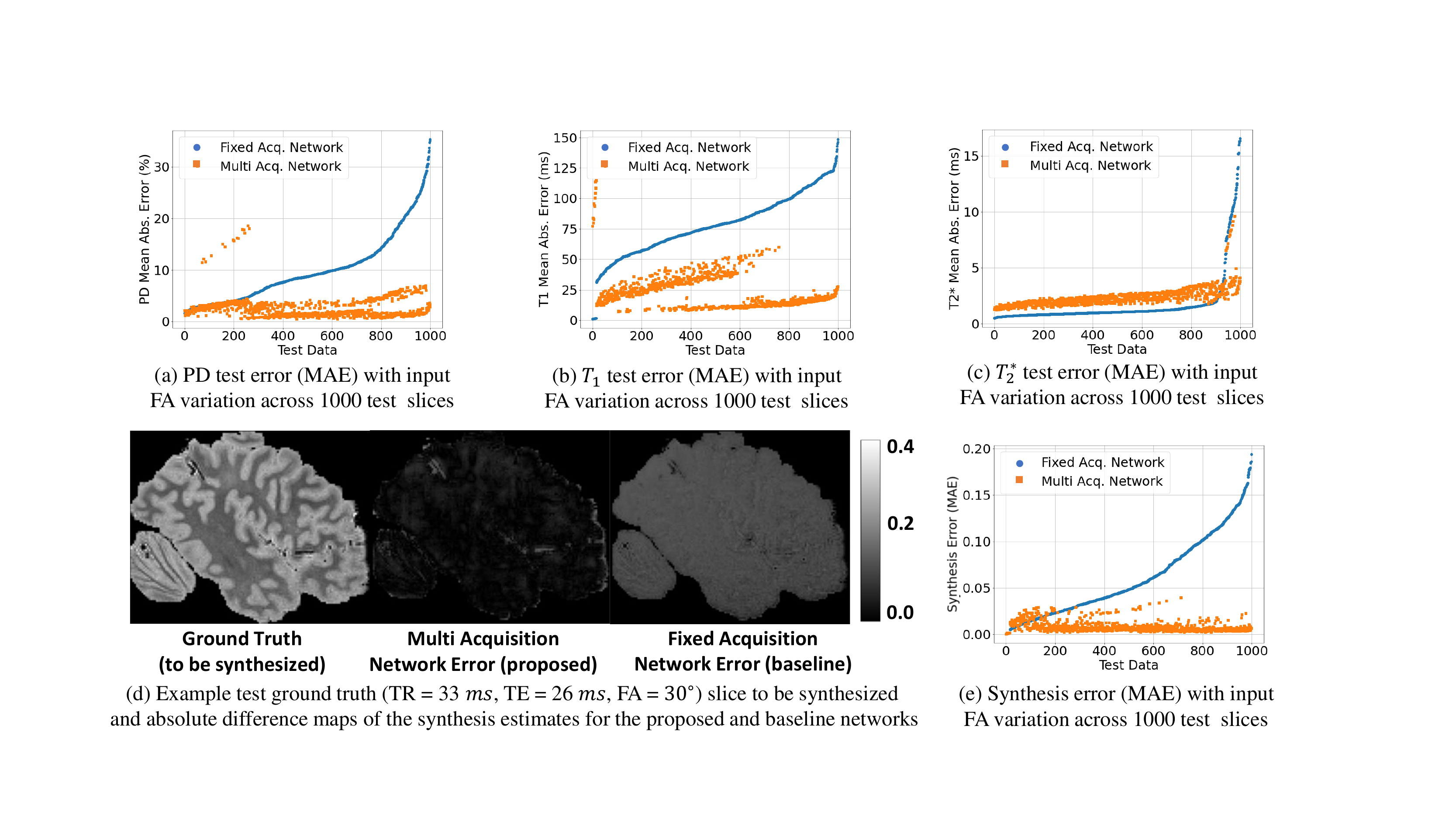}
\caption{Scan Parameter Robustness. Test MAE in synthesis with input acquisition parameter ($\mathbf{\phi}_i$) variation. The proposed network is more robust to input variation because of being exposed to scans acquired with random parameters. Panel d show the ground truth and synthesis normalized absolute difference error maps of a single test image slice for which the flip angle had been perturbed. \vspace{-0.5cm} }
 \label{fig:acq_perturb} 
\end{figure*}

\begin{figure}[h!]
\centering
\includegraphics[trim=120 100 200 80, clip, width=0.8\textwidth]{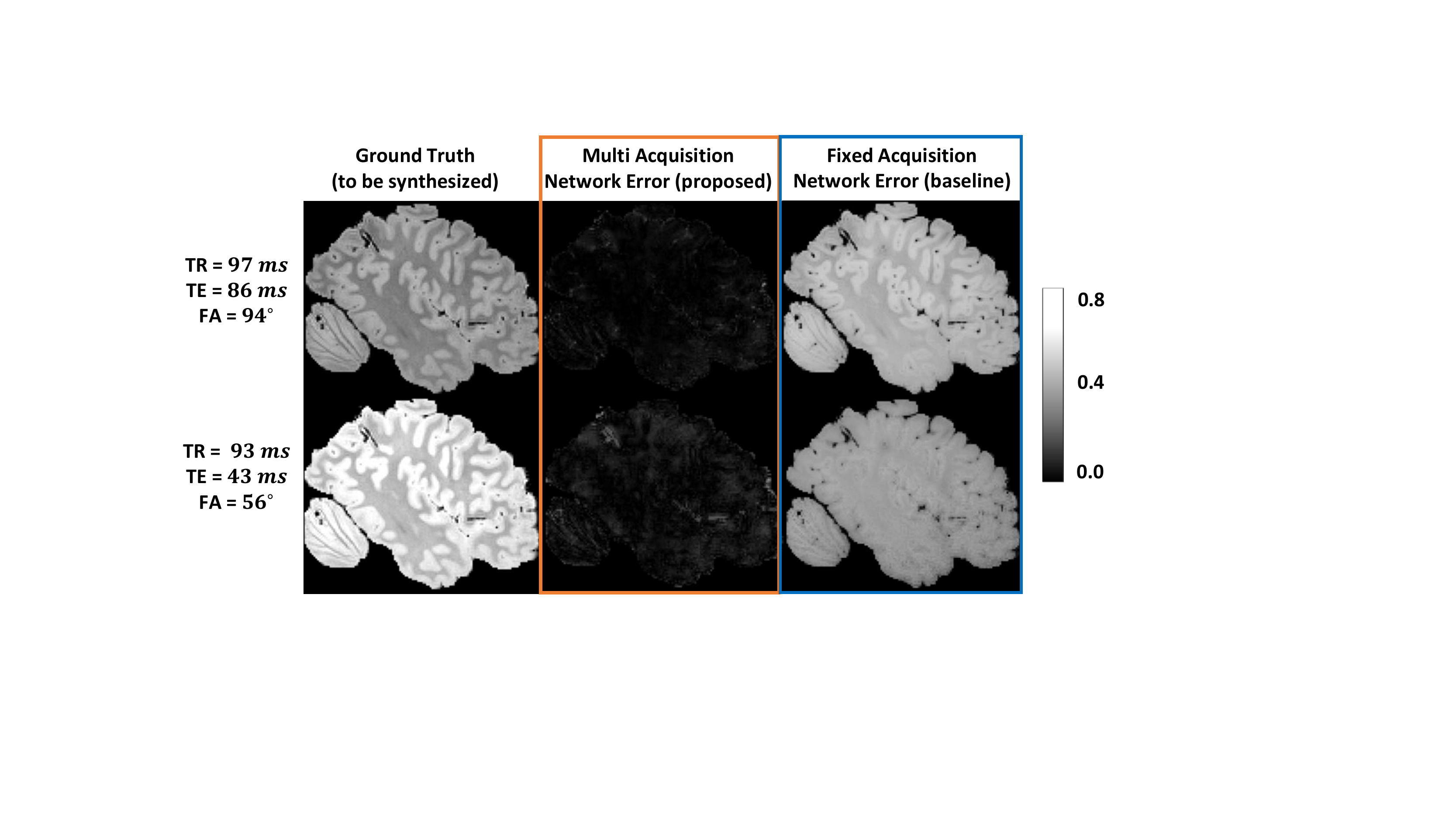}
\vspace{-0.2cm}
\caption{Additional example for testing generalizability to varying scan parameters (Experiment 2). Normalized absolute difference error maps of MRI image synthesis estimates for which the flip angle had been perturbed. The proposed network substantially outperforms the baseline network in synthesizing unseen MRI contrasts. \vspace{-0.5cm}}
\label{fig:acq_perturb2} 
\end{figure}

We evaluated the performance of the proposed method to estimate tissue parameters and subsequently synthesize new contrasts with test data acquired with varying flip angles. Since existing methods are trained for a given acquisition protocol and cannot be used in clinical sites that use different scanner parameters, we hypothesize that the proposed network will outperform the baseline in estimating parameters that depend on the flip angle scanner parameter, i.e. $\mathbf{T_1}$ and $\mathbf{PD}$, and will overall perform better in synthesizing. 
We randomly perturbed the flip angle by $\pm$ (0 - 20) degrees to simulate heterogeneous data with varying input scanner parameters. We analyzed on parameter estimation and synthesis accuracy over 1000 test slices to evaluate the methods.

\noindent \textbf{Results}.
Fig.~\ref{fig:acq_perturb} reports the generalizability performance of the proposed approach to perturbations in the input flip angle acquisition parameter $\scparin$. The proposed method yields either comparable or lower errors for 98.4\% cases for (a) $\mathbf{T_1}$ estimation, (b) $PD$ estimation, and (e) synthesis across 1000 different testing configurations. The estimation error of $\mathbf{T_2^*}$ is comparable for both methods for 900 test cases. However, for 100 test cases which largely comprise of flip angle configurations that were close to 20 degrees different from the angle used in training the baseline network, the performance of the proposed method remains consistent, while the baseline error is substantially higher. These results demonstrate that the proposed method accurately estimates all tissue properties and synthesizes from contrasts of multiple acquisition protocols across heterogeneous datasets. Fig.~\ref{fig:acq_perturb}d illustrates an example ground truth scan and absolute difference maps for the synthesis estimates of the ground truth. The substantial improvement in synthesis errors provided by the proposed method is directly due to improvements in the tissue parameter estimates. Our results indicate that varying the acquisition settings in the training data resulted in substantial improvements, especially in cases where the baseline errors deteriorated,  suggesting that generalizable networks that can be used across multiple acquisition protocols can be deployed in practice without much loss in performance. Additional examples in Fig.~\ref{fig:acq_perturb2} are in agreement with our analysis demonstrating the superior generalizability of the proposed method compared to the baseline.

\vspace{-0.1cm}
\subsection{Experiment 3: MRI synthesis}
\vspace{-0.1cm}

\def\arraystretch{0.5}%
\begin{figure*}[t]
\centering
\includegraphics[trim=50 310 100 20, clip, width=0.95\textwidth]{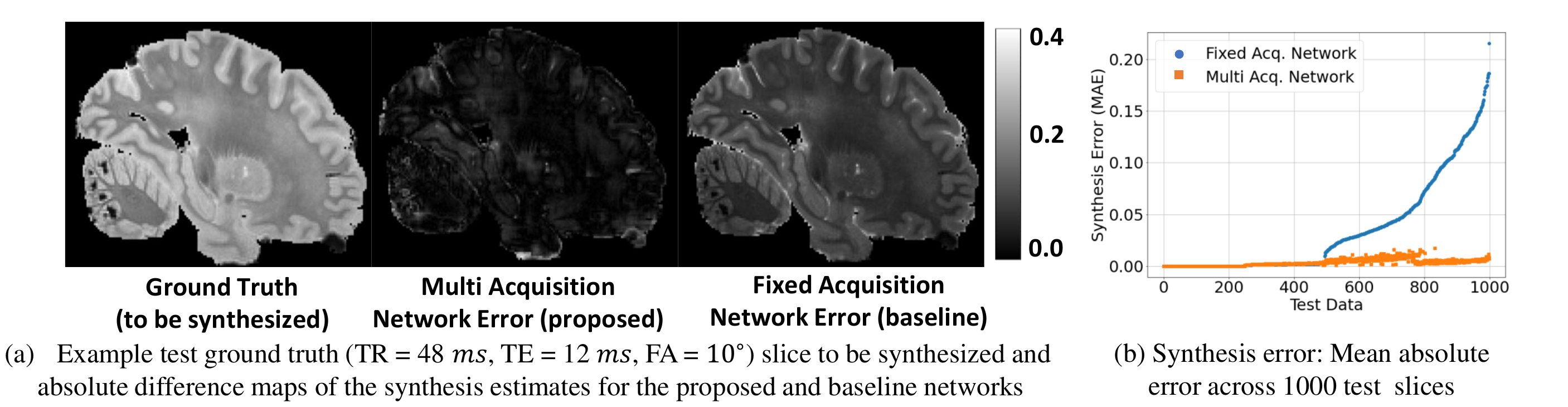}
\vspace{-0.2cm}
\caption{Contrast Synthesis. Test error in synthesis of image contrasts estimated from single scan 3-echo MRI over 500 test images. Subfigures b and c show the normalized absolute difference error for an example synthesized test image using the proposed and baseline methods, respectively. \vspace{-0.5cm}}
\label{fig:mse_synth} 
\end{figure*}

We evaluated the ability of the proposed method to synthesize arbitrary unseen FLASH MRI contrasts from a single multiecho input MRI scan session. 
We fixed the scanner parameters $\scparin$ of the test data to those used for training the baseline for all the 1000 test slices, the optimal scenario for the baseline method. We randomly generated the scanner parameters $\scparout$ of the output/synthesized scans. Details of scan parameter values is provided in Table 2. We compared the MAE of the proposed and the baseline method with the ground truth for the 1000 test slices to evaluate synthesis accuracy. 

\textbf{Results.}
Fig~\ref{fig:mse_synth} compares the synthesis accuracy of the proposed method with the baseline. The baseline reconstruction for an example slice shown in Fig~\ref{fig:mse_synth}a. has large errors while the proposed method exhibits substantially lower values in the error image. The MAE plot in Fig.~\ref{fig:mse_synth}b. demonstrates that the proposed method achieves comparable or substantially lower error across all  1000 test slices.

\subsection{Experiment 4: Effect of forward model based synthesis loss}
~\label{sec:ab1}
\vspace{-0.2cm}

\begin{figure}[t]
\centering
\includegraphics[trim=90 250 60 80, clip,width=0.95\textwidth]{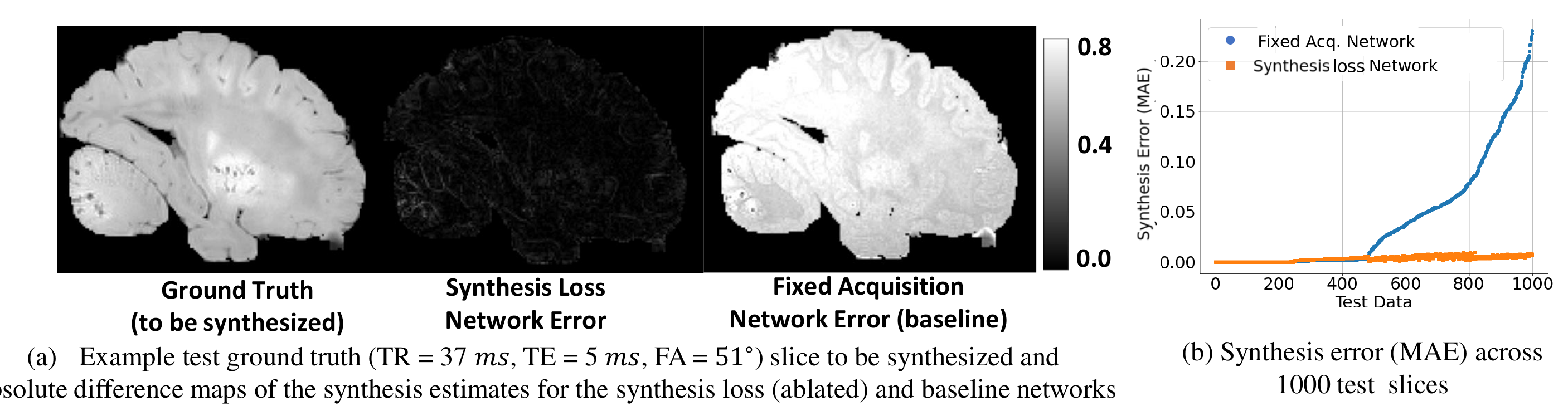}
\vspace{-0.1cm}
\caption{Mean absolute error measured across 1000 test images for  networks. \vspace{-0.3cm}}
 \label{fig:seploss} 
\end{figure}

Existing learning based tissue property estimation methods are optimized for specific input scans they are trained on, and optimizing the (unseen) synthesis scans has not been demonstrated before. We conducted an ablation study to understand the benefits of separating the scanner parameters~$\scparin$ of input contrasts from those used in the loss function~($\scparout$). We compare the synthesis performance of the proposed and baseline fixed acquisition network, which only differ in the contrasts that they optimize. Table 1. shows training scan parameters and Table 2 shows test scan parameters.

\textbf{Results.}
Fig.~\ref{fig:seploss}a. shows an example ground truth of a test slice and the corresponding absolute difference maps with the synthesis estimations from both methods. The synthesis loss network has lower errors than the baseline fixed acquisition network. Fig.~\ref{fig:seploss}b. compares the synthesis test error across 1000 test slices. 
The proposed synthesis loss strategy consistently outperforms the baseline network that optimizes for input contrasts with the maximum error reported to be 200 times more for the latter. Separating the input contrast from the contrast used as target increases the overall synthesis accuracy of the method. This is an important result, as existing work in MRI parameter estimation literature use the same input contrast as input and target output, thereby limiting the learning capability of the network.

\section{Discussion and Conclusion}
We present a novel unsupervised learning method to estimate tissue properties and synthesize arbitrary FLASH MRI contrast from a single multiecho scan session. The proposed strategy involves training with multiple acquisition protocols, leading to generalizations across a variety of acquisitions, and optimized FLASH contrasts that were different than the input contrast. Our results demonstrated the improved performance and generalizability of this approach over ablated methods that are representative of  baseline learning strategies. While we focus on the medical imaging application, similar unsupervised synthesis networks are common in other domains where we believe our analysis could potentially generalize to them.

Our method could directly impact patient diagnosis as estimated tissue properties can be used by to make clinical decisions. Incorrect estimations could lead to incorrect clinical decisions, negatively impacting patients. The black box nature of the solution makes it challenging to predict conditions under which these estimations can fail. Care must be taken to combine additional information, including scans that are directly acquired and minimally processed, to ensure reliability across imaging data. Training large networks also yield negative environmental impacts. The proposed solution which uses 2D training and broadly generalizable networks that do not require multiple training rounds help mitigate this negative impact.

We use a least squares cost function where the inherent assumption is that the measurement noise model is Gaussian distributed. However, the noise model can become inaccurate for low SNR magnitude MRI, leading to an estimation bias in the parameters~\cite{sijbers2004, aja-fernandez2008, varadarajan2015}. Many approaches in the literature the full complex valued MRI where the noise is Gaussian distributed~\cite{liu2019,cai2018}. However,for complex-valued methods to be incorporated into routine clinical imaging in hospitals will require changes to the scanner software and saving double the data - both will require a long term plan and a change in clinical culture. In addition, these methods will not work with existing datasets. Hence, a future direction for this work will be to  extend the optimization to minimize the likelihood of the appropriate noise model. 

Additional system imperfections can be incorporated into the FLASH model~\cite{balbastre2020}, leading to estimations robust to intensity and contrast variations, such as the spatial variation of the flip angle due to dielectric effects caused by a non-uniform radio frequency (RF) field generated by the RF transmit coil used in the acquisition. Our result in Fig~\ref{fig:acq_perturb} showcased the ability of the proposed method to handle variation in flip angles successfully. Building on this preliminary result, incorporating the spatial variation in the forward model and as data augmentation during training is a natural extension to further improve robustness of the proposed approach.

In this paper we have focused on synthesizing new FLASH MRI contrasts from estimated properties. However, there are several other imaging contrasts that can be generated from the tissue property estimates. These contrast use different sequences (or scanner parameter encoding) and hence have different forward models. While most forward models cannot be analytically represented, they can be simulated by an iterative mechanism using Bloch equations. Extending this work to other MRI sequences could potentially enable us to improve the overall accuracy of our parameter estimates and further increase the generalizability of the proposed approach to take other MRI contrasts as input.


{
\small
\bibliographystyle{unsrt}
\bibliography{references}
}

\end{document}